\documentclass[aps,prd,twocolumn, superscriptaddress,showpacs,preprintnumbers,amssymb]{revtex4-1}

\usepackage{color}
\usepackage{graphicx} % Include figure files

\usepackage{multirow}
\usepackage{array}
\usepackage{pict2e}
\usepackage{hyperref}
\graphicspath{{ps}}
\usepackage{adjustbox}
\usepackage{rotating}
\usepackage{siunitx}
\usepackage{orcidlink}
\usepackage[fleqn]{amsmath}
\usepackage{siunitx}
\newcolumntype{T}[1]{S[table-format=#1]}

\begin{document}

\title{\quad\\[0.5cm]Search for baryon and lepton number violating decays $\boldsymbol{D \rightarrow p\ell}$}

%%% Paper:    D to p l
%%% Journal:  Physical Review D Letter
%%% Contacts: S. Maity, S. Bahinipati, V. Bhardwaj, R. Garg
%%% Non-responding authors or those who said NO are commented out.
%%% ====================================================================
%%% Click the RELOAD button on your web browser to see the updated file.
%%% ====================================================================
%%% Use \input{pub658-orcid} to insert this material into your latex file.
\noaffiliation
% \author{I.~Adachi\,\orcidlink{0000-0003-2287-0173}} % 2590
% \author{K.~Adamczyk\,\orcidlink{0000-0001-6208-0876}} % 2239
% \author{J.~K.~Ahn\,\orcidlink{0000-0002-5795-2243}} % 7423
  \author{S.~Maity\,\orcidlink{0000-0003-3076-9243}} % 2985
  \author{R.~Garg\,\orcidlink{0000-0002-7406-4707}} % 2213
  \author{S.~Bahinipati\,\orcidlink{0000-0002-3744-5332}} % 2332 
  \author{V.~Bhardwaj\,\orcidlink{0000-0001-8857-8621}} % 2228
  \author{H.~Aihara\,\orcidlink{0000-0002-1907-5964}} % 2223
  \author{S.~Al~Said\,\orcidlink{0000-0002-4895-3869}} % 6823
  \author{D.~M.~Asner\,\orcidlink{0000-0002-1586-5790}} % 4684
  \author{H.~Atmacan\,\orcidlink{0000-0003-2435-501X}} % 2538
% \author{V.~Aulchenko\,\orcidlink{0000-0002-5394-4406}} % 8183
  \author{T.~Aushev\,\orcidlink{0000-0002-6347-7055}} % 3747
  \author{R.~Ayad\,\orcidlink{0000-0003-3466-9290}} % 3766
% \author{T.~Aziz\,\orcidlink{-}} % 3523
% ...
% 2387
% \author{V.~Zhulanov\,\orcidlink{0000-0002-0306-9199}} % 4983
\collaboration{The Belle Collaboration}

\begin{abstract}
We search for the baryon and lepton number violating charm decays, $D \rightarrow p\ell$, where $D$ is either a $D^0$ or a $\overline{D}^0$ and $\ell$ is a muon or an electron, using a data sample of $921\,\mathrm{fb}^{-1}$ collected by the Belle detector at the KEKB asymmetric energy $e^{+}e^{-}$ collider. %We observe event yields with significance less than 1.6$\sigma$ for any of the decay modes. 
In the absence of significant signals, we set upper limits on the branching fractions in the range $(5 - 8)  \times 10^{-7}$ at a 90\% confidence level, depending on the decay mode.
\end{abstract}
\pacs{13.25.Ft, 14.65.Dw, 14.80.Sv, 13.20.Fc, 13.66.Bc, 14.65.Dw}
\maketitle
\begin{figure}[!h]
	
        \includegraphics[ width=0.20\textwidth]{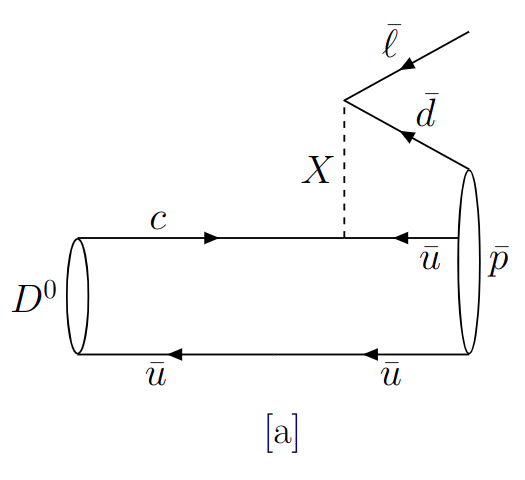}
        \includegraphics[ width=0.20\textwidth]{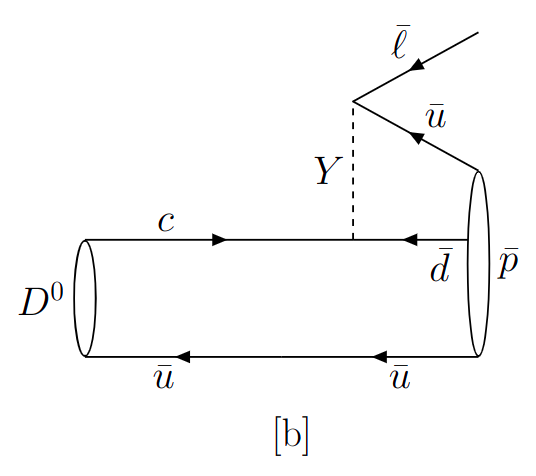}
       
 \caption{Feynman diagrams of the decays $D^{0} \rightarrow \overline{p}\ell^{+}$ with non-SM gauge bosons (a) $X$ and (b) $Y$.}
	
	\label{fig:lepto}
\end{figure}

 
Baryon number violation (BNV) is one of the crucial ingredients to create the matter-antimatter asymmetry as observed in the universe~\cite{Sakharov}. The known particles and antiparticles and the interactions among them are described by the Standard Model (SM). 
Several grand unified theories (GUTs)~\cite{Pati, Georgi, Raby, Davidson, Mohapatra}, supersymmetry and other SM extensions~\cite{Lola, Polonsky} propose BNV processes of nucleons. In these models, baryon ($B$) and lepton ($L$) numbers are explicitly violated but the difference between these numbers is conserved i.e., $\Delta (B-L)=0$.
%($\Delta \lvert B-L\rvert=0$). 
Several attempts ~\cite{Nishino, Shiozawa, Hayato, Antonio} have been made to search for the decays of the lightest baryon, namely the proton, but no evidence for its decay has yet been found. Searches for decays of heavy mesons to final states with nonzero $B$ values can provide an alternative probe for BNV decays. 
Various non-SM models of proton decays can be extended to predict possible decay mechanisms for $D$ decays. For example, analogous to proton decays~\cite{Biswal}, the decays of $D^{0} \to \overline{p} \ell^{+}$ can be explained using leptoquark couplings. Figure~\ref{fig:lepto} shows possible Feynman diagrams for $D^{0} \rightarrow \overline{p}\ell^{+}$ decays, where the mediators are non-SM gauge bosons. Decays of the $D$ meson to final states containing a proton probe BNV within the first two generations.

The $D \rightarrow p\ell$ decays simultaneously violate $B$ and $L$ but conserve $(B-L)$. Further, leptoquarks have been proposed to explain recent anomalies reported by different experiments~\cite{Wmass, RD, muong2} and the $D \rightarrow p\ell$ search will provide valuable input to the search for leptoquarks.

Experimentally, various BNV processes in $D$, $B$, and $\Lambda$ decays were searched for by the CLEO~\cite{Rubin}, B{\small{A}}B{\small{AR}}~\cite{Sanchez}, and CLAS ~\cite{McCracken} collaborations, respectively, but no evidence for a signal was found. The BESIII collaboration has searched for $D \to p e$ decay modes and set 90\% confidence-level upper limits on the branching fractions of $\mathcal{B}(D^0 \to \overline{p} e^+) < 1.2 \times 10^{-6}$ and $\mathcal{B}(D^0  \to p e^-) < 2.2 \times 10^{-6}$~\cite{Ablikim}, and these are currently the most stringent limits. The large data sample collected by the Belle experiment provides improved sensitivity for BNV decays of charm mesons. We set limits for the $D^{0}$ and $\overline{D}^{0}$ decays separately for the first time.  
In this paper, we report a search for the $D$ meson decay modes $D^{0} \rightarrow p\ell^{-}$, $\overline{D}^{0} \rightarrow p\ell^{-}$, $D^{0} \rightarrow \overline{p}\ell^{+}$, and $\overline{D}^{0} \rightarrow \overline{p}\ell^{+}$, where $\ell$ is $e$ or $\mu$, using the data collected with the Belle detector at the KEKB asymmetric-energy $e^+e^-$ collider~\cite{kekb1,kekb2} located at the High Energy Accelerator Research Organisation in Japan. The data used in this analysis were collected at $e^+e^-$ center-of-mass (c.m.) energies at and 60 MeV below the $\Upsilon$(4S) resonance, and at the $\Upsilon$(5S) resonance with integrated luminosities of 711 fb$^{-1}$, 89 fb$^{-1}$, and 121 fb$^{-1}$, respectively. The total integrated luminosity is 921 fb$^{-1}$. Throughout this paper, charge conjugate modes are implicitly included, unless otherwise noted.

The Belle detector is a large solid-angle magnetic spectrometer that consists of a silicon vertex detector (SVD), a 50-layer central drift chamber (CDC), an array of aerogel threshold Cherenkov counters (ACC), a barrel-like arrangement of time-of-flight (TOF) scintillation counters, and a CsI(Tl) crystal based electromagnetic calorimeter (ECL); all are located inside a superconducting solenoid coil that provides a 1.5 T magnetic field. An iron flux-return yoke placed outside the coil is instrumented with resistive plate chambers to detect $K_{L}^{0}$ mesons and muons (KLM). A more detailed description of the detector can be found in Ref.~\cite{Belle:2000cnh}. 

Monte Carlo (MC) simulation samples are used to optimize the selection criteria, estimate the signal reconstruction efficiencies, and identify and model the distributions of various sources of background. The signal MC sample for each decay mode is generated using EvtGen~\cite{evtgen} where $e^{+}e^{-}\rightarrow q\overline{q}$ generated with Pythia~\cite{pythia} and PHOTOS~\cite{photos} takes into account final state radiations. The produced particles are propagated with GEANT3~\cite{geant3} to simulate a detector response. The well-established $D^0 \to K^- \pi^+$ decay mode is used for normalization to measure the branching fractions of signal modes.

Reconstructed trajectories of charged particles (tracks) are required to originate from the interaction point (IP) and have a point of closest approach to the latter within 3.0 cm along the $e^{+}$ beam axis and 1.0 cm in the transverse plane. These requirements remove tracks not originating from the IP. Additionally, the tracks are required to have at least two hits in the SVD. 
The final-state charged hadrons (pions, kaons, and protons) are identified based on the number of Cherenkov photons in the ACC, and TOF and $dE/dx$ measurements. All of this information is combined to form pion, kaon, and proton likelihoods,  $\mathcal{L}_{\pi}, ~\mathcal{L}_{K}$, and $\mathcal{L}_{p}$, respectively. The selection is made on the basis of likelihood ratios, $\mathcal{L}_{i/j} = \mathcal{L}_{i}/(\mathcal{L}_{i} + \mathcal{L}_{j})$, where $i$ and $j$ are $\pi, K$ or $p$. Protons are identified by requiring $\mathcal{L}_{p/\pi} > 0.6$ and $\mathcal{L}_{p/K} > 0.6$ with an identification efficiency of 90\% while the probability of misidentifying a kaon or a pion as a proton is 3\%. Kaons and pions are selected by requiring  $\mathcal{L}_{K/\pi} > 0.6$ and $\mathcal{L}_{\pi/K} > 0.6$, respectively. The identification efficiencies for kaons and pions are 95\% and 94\%, respectively. The probability of misidentifying a pion (kaon) as a kaon (pion) is 4\% (5\%). 

The electromagnetic shower shape, $E_{\rm ECL}/p$ ratio, where $E_{\rm ECL}$ is the energy deposition in the ECL and $p$ is the track momentum, and the position matching between track and ECL cluster are utilized for electron identification, in addition to the information used for charged hadron identification excluding that from the TOF. All this information is combined to form an electron likelihood ratio, $\mathcal{L}_{e}$ and the electrons are identified requiring $\mathcal{L}_{e} > $ 0.9. 
To recover the energy loss due to bremsstrahlung, we search for photons in a cone of 50\,mrad around the initial direction of the electron momentum; if found, their momenta are added to that of the electron. Muons are identified by using the track penetration depth, hit distribution pattern in the KLM, and matching quality. This information is combined to form a muon likelihood $\mathcal{L_{\mu}}$. Muons are identified by requiring a likelihood ratio of $\mathcal{L_{\mu}}/ (\mathcal{L_{\mu}} + \mathcal{L_{\pi}} + \mathcal{L}_{K}) > 0.9$. The electron (muon) identification efficiency for these criteria is 93\% (94\%) with a probability of misidentifying a pion as an electron (a muon) below 0.5\% (4\%). The kaon-to-electron misidentification rate is negligible, while the probability of identifying a kaon as a muon is similar to that of identifying a pion as a muon. All the quoted identification efficiencies and misidentification probabilities are averaged over the momenta of final-state particles. 
 
Tracks with momentum greater than 0.6 GeV/$c$ are used to reconstruct $D^0$ candidates in the $K^-\pi^+$, $pe^-$, $\overline{p} e^+$, $p \mu^-$, and $\overline{p}\mu^+$ final states. The invariant mass of a $D^0$ candidate is required to be in the range 1.8 $ < M_{D^0} < $ 1.9 GeV/$c^2$. We reconstruct a $D^{*+}$ candidate by combining the $D^0$ candidate with a slow pion ($\pi^+_{s}$). We require all the signal $D^{0}$s to come from $D^{*+}$, where $D^{0}$ and $\overline{D}^{0}$ flavors are determined by the $\pi_{s}$ charge in $D^{*+} \rightarrow D^{0}\pi^{+}$.
The mass difference, $\Delta M = M_{D^{*+}}-M_{D^0}$, where $M_{D^{*+}}$ ($M_{D^0}$) is the invariant mass of the $D^{*+}$ ($D^0$) candidate, is required to be less than 0.158 GeV$/c^{2}$. In order to improve the $\Delta M$ resolution, the $\pi^+_{s}$ is constrained to originate from the IP. The momentum of the $D^{*+}$ candidate in the c.m.\ frame is required to be greater than 2.5\,GeV$/c$ to reduce $B\overline{B}$ and combinatorial backgrounds. 
A vertex fit is performed on the selected $D^{*+}$ candidates. After applying all the selection criteria, a small fraction (0.4\%) of the selected events have more than one $D^{*+}$ candidate. In an event with multiple $D^{*+}$ candidates, we select the one that has the lowest $\chi^{2}$ value associated with the $D^{*+}$ vertex fit. The efficiency of such candidate selection ranges from 85 to 94\% depending on the decay mode and is 83\% for the normalization mode.

We study the background using simulated samples corresponding to an integrated luminosity of about 4.5 times that of the Belle data sample. In the $\overline{D}^{0} \rightarrow pe^{-}$ and $D^{0} \rightarrow \overline{p}e^{+}$ modes, background from semileptonic $D \rightarrow Ke\nu_{e}$ decay is found to peak in the $\Delta M$ signal region, while for the $D^{0} \rightarrow pe^{-}$ and $\overline{D}^{0} \rightarrow \overline{p}e^{+}$ decay modes, no such peaking structure is observed. Peaking background in the $\Delta M$ distribution of the $\overline{D}^{0} \rightarrow p\mu^{-}$ and $D^{0} \rightarrow \overline{p}\mu^{+}$ modes arises from $D \rightarrow K\rho$, $K^{*}\pi$, $KK$, and $K\mu\nu_{\mu}$ decays. In $D^{0} \rightarrow p\mu^{-}$, a tiny peaking structure is observed from the $D \rightarrow KK$ decay mode. The peaking structure in the aforementioned modes is observed owing to the misidentification of final-state particles. There is no peaking structure in $M_{D^{0}}$ for any of the signal modes. 

For the $D^{0} \rightarrow K^{-}\pi^{+}$ normalization mode, small background contributions originate from the correct $D^{0}$ candidate's combination with a random $\pi_{s}$ as well as from partially reconstructed $D^{0}$ decays. The signal purity for  $D^{0} \rightarrow K^{-}\pi^{+}$ is 95\% in a region around the $M_{D^{0}}$ and $\Delta M$ signal peaks that contain 78\% of the $D^{0}$ candidates. We reconstruct over 1.7 million events with an efficiency of 13.5\%.

Signal yields are extracted with extended maximum-likelihood fits to the unbinned $M_{D^{0}}$ and $\Delta M$ distributions of each decay mode. Separate probability density functions (PDFs) are used for signal, peaking and combinatorial backgrounds. For signal events, the PDF for the $\Delta M$ distribution is parameterized by using the sum of four Gaussians, while the $M_{D^{0}}$ PDF is modeled by the sum of two Gaussians and one asymmetric Gaussian with separate parameters for each decay mode. The core Gaussian parameters for the $D^0 \to K^{-}\pi^{+}$ decay mode are allowed to vary in the fit, while for the $D^0 \to p \ell$ decay modes, they are fixed to the values obtained from simulated signal events calibrated with data; the calibration is determined by comparing the shape parameters between data and simulation for the $D^0 \to K^{-}\pi^{+}$ decay. All the remaining shape parameters of signal PDF are fixed to those obtained in the fit to simulated signal events. 

The peaking background PDF in $\Delta M$ is modeled by the sum of two asymmetric Gaussians for the $\overline{D}^{0} \rightarrow pe^{-}$ and $D^{0} \rightarrow \overline{p}e^{+}$ modes. The sum of the $D \rightarrow K\mu\nu_{\mu}$, $D \rightarrow K\rho$ and $D \rightarrow K^{*}\pi$ components is modeled by the sum of two asymmetric Gaussians for the $\overline{D}^{0} \rightarrow p\mu^{-}$ and $D^{0} \rightarrow \overline{p}\mu^{+}$ modes. The $D \rightarrow KK$ component is parameterized by the sum of two Gaussians for the $D^0 \to p\mu^{-}$ decay mode. The background components in $M_{D^{0}}$ are modeled by a first-order polynomial for all the decay modes. 
For the normalization mode, the background from random $\pi_{s}$ in $\Delta M$ is parameterized by a threshold function, defined as
\begin{equation}
    f(x) = (\Delta M-m_\pi)^{a} e^{-b(\Delta M-{m_{\pi}})},
\end{equation}
where $m_{\pi}$ is the known charged pion mass~\cite{pdg} and $a$, $b$ are the shape parameters. The background from partially reconstructed $D$ decays is parameterized by a Gaussian in $\Delta M$. The aforementioned two backgrounds are parameterized by a Gaussian function and a second-order polynomial, respectively, in $M_{D^0}$.

\begin{figure}[!h]
        \includegraphics[trim = {0 3.5cm 0 0}, width=0.50\textwidth]{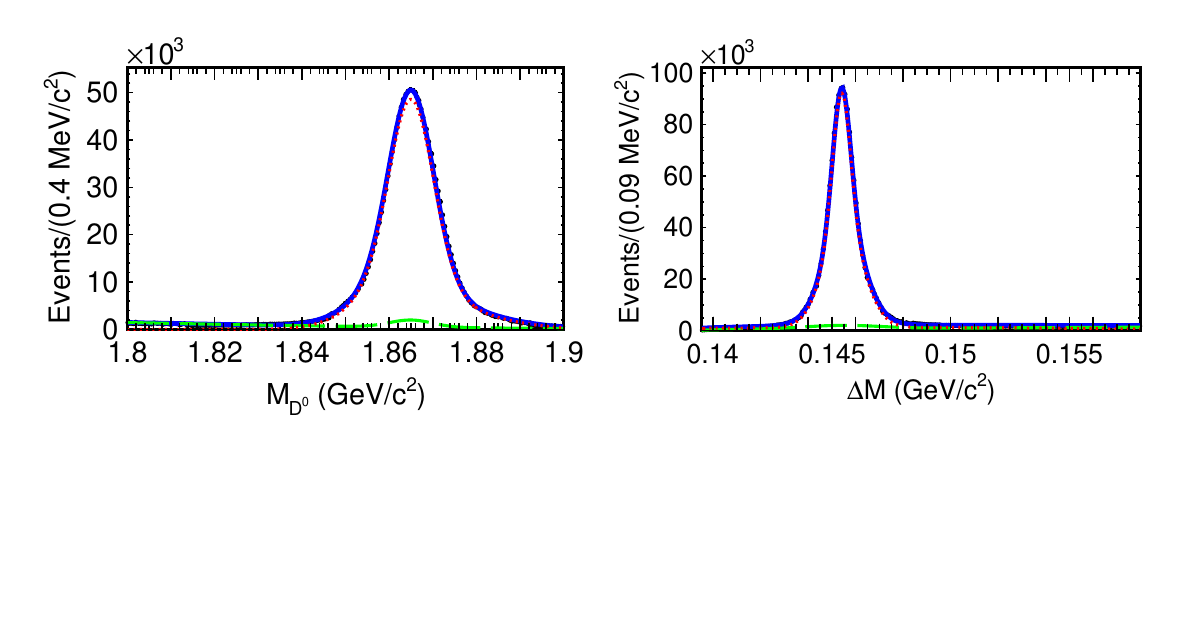}
        
	\caption{Fit projections for the $M_{D^0}$ (left) and $\Delta M$ (right) distributions for the $D^0 \to K^- \pi^+$ normalization mode. The red dotted curves show the fit function for the signal, the green dashed curves show the fit function for the total background, and the blue solid curves show the sum of the fit functions.}
	%\caption{(a) and (b) are the possible decay diagrams of $D^{0} \rightarrow \overline{p}\ell^{+}$.}
	\label{figur:fit_results_kpi}
\end{figure}

\begin{figure}[!h]
	 \includegraphics[width=0.48\textwidth]{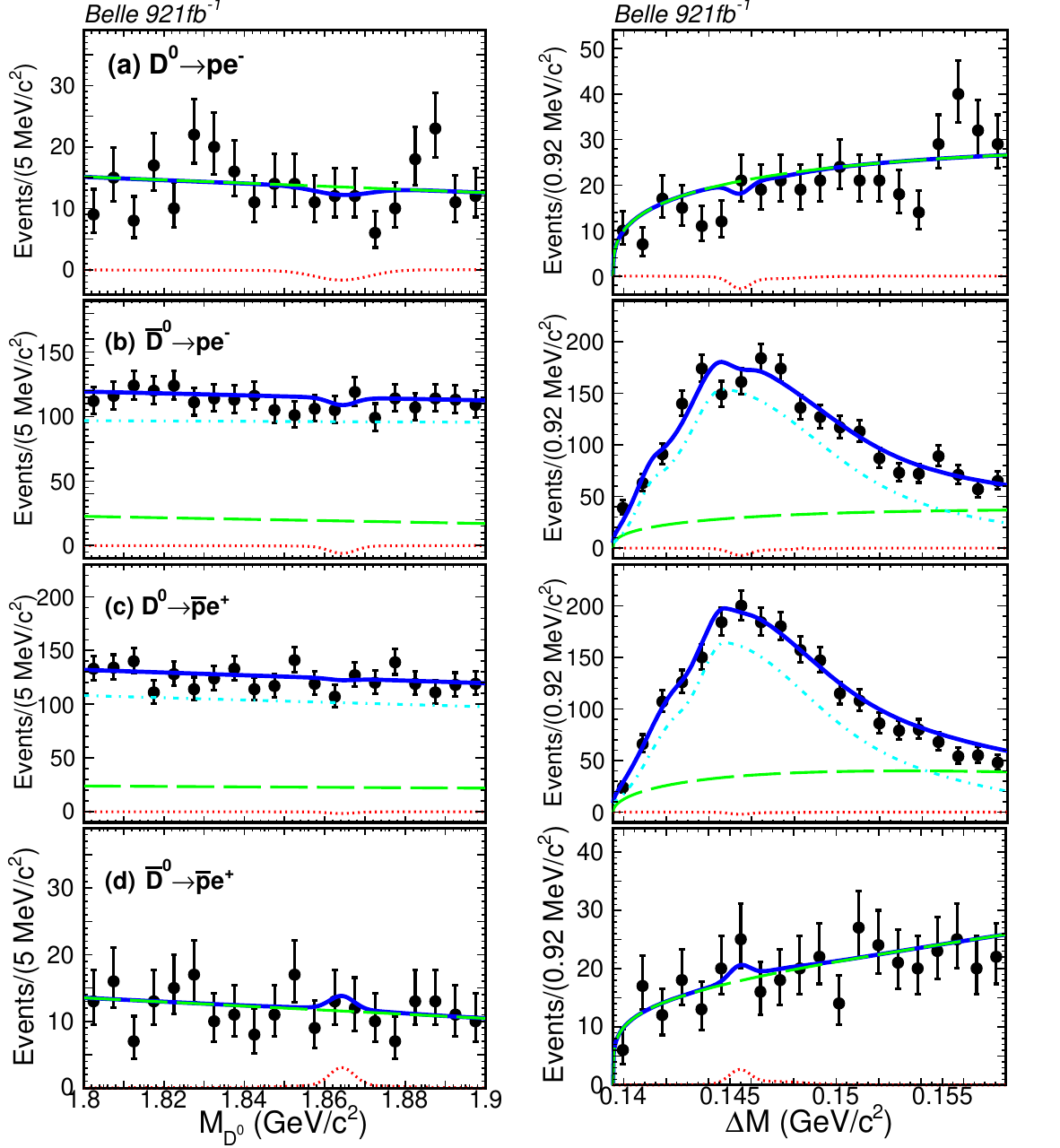}
 \caption{Fit projections for the $M_{D^0}$ (left) and $\Delta M$ (right) distributions for the (a) $D^0 \to p e^-$, (b) $\overline{D}^0 \to p e^-$, (c) $D^0 \to \overline{p} e^+$, and (d) $\overline{D}^0 \to \overline{p} e^+$ decay modes. The red dotted curves show the fit function for the signal, the green dashed curves show the fit function for the combinatorial background, the cyan dashed-dotted curves show the fit function for the peaking background, and the blue solid curves show the sum of the fit functions.}
	\label{figur:fit_results_data_e}
\end{figure}
\begin{figure}[!h]
 \includegraphics[ width=0.48\textwidth]{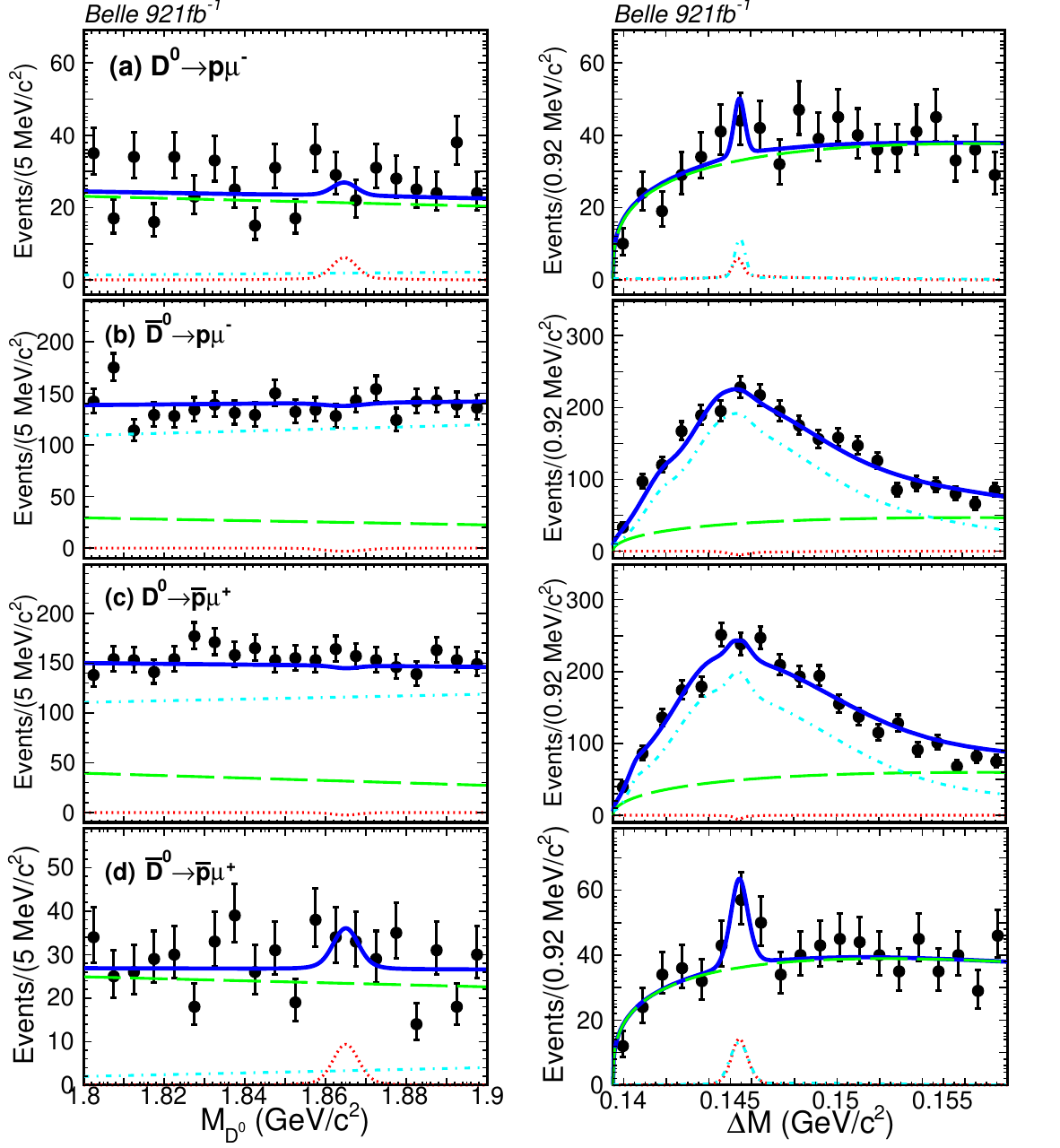}
	\caption{Fit projection for the $M_{D^0}$ (left) and $\Delta M$ (right) distributions for the (a) $D^0 \to p \mu^-$, (b) $\overline{D}^0 \to p \mu^-$, (c) $D^0 \to \overline{p} \mu^+$, and (d) $\overline{D}^0 \to \overline{p} \mu^+$ decay modes. The red dotted curves show the fit function for the signal, the green dashed curves show the fit function for the combinatorial background, the cyan dashed-dotted curves show the fit function for the peaking background, and the blue solid curves show the sum of the fit functions.}
	\label{figur:fit_results_data_mu}
\end{figure}

For the combinatorial background for each decay mode, the PDF for $\Delta M$ is parameterized by the threshold function. 
\noindent
%The $\Delta M$ distribution is fitted from the $\pi^+$ mass to 0.158 GeV/c$^{2}$.
The combinatorial background PDF for $M_{D^0}$ is modeled by a first- and second-order polynomial for the signal and normalization decay mode, respectively. The parameters of the combinatorial background PDF for signal modes are fixed except for $D^{0} \rightarrow pe^{-}$, $\overline{D}^{0} \rightarrow \overline{p}e^{+}$, $D^{0} \rightarrow p\mu^{-}$, and $\overline{D}^{0} \rightarrow \overline{p}\mu^{+}$ decay modes in which cases they are floated. The parameters of the combinatorial background PDF for the $D^{0} \rightarrow K^{-}\pi^{+}$ mode are floated.
The fit projections for the $M_{D^{0}}$ and $\Delta M$ distributions are shown in Figs.~\ref{figur:fit_results_kpi}, ~\ref{figur:fit_results_data_e} and~\ref{figur:fit_results_data_mu} for the $D^{0} \rightarrow K^{-}\pi^{+}$, $D \rightarrow p e$, and $D \rightarrow p \mu$ decay modes, respectively. The obtained signal yields are listed in Table~\ref{tab:ul_data}. 

 \begin{table}[htbp]
   
	\begin{center}	
	%\label{tab:ul_data}
 \caption{\label{tab:ul_data} Reconstruction efficiency ($\epsilon$), signal yield ($N_{S}$), signal significance ($\mathcal{S}$), upper limit on the signal yield ($N_{p\ell}^{UL}$), and branching fraction ($\mathcal{B}$) at 90\% confidence level for each decay mode.}

	\resizebox{8cm}{!}{%
		\begin{tabular}{c c c c c c}
			\hline \hline
			Decay mode & $\epsilon$ (\%)& $N_S$ &$\mathcal{S}$ ($\sigma$)&$N_{pl}^{UL}$ & $\mathcal{B} \times 10^{-7}$ \\ [0.5ex] 
			\hline \hline
			$D^{0} \rightarrow pe^{-}$ & 10.2 &\phantom{0} $-6.4 \pm 8.5$ & \textemdash& 17.5 & $ < 5.5 $  \\  [1ex]
			%\hline
			$\overline{D}^{0} \rightarrow pe^{-}$ & 10.2 & \phantom{0} $-18.4 \pm 23.0$ & \textemdash& 22.0 & $< 6.9 $  \\  [1ex]
			%\hline
			$D^{0} \rightarrow \overline{p}e^{+}$ & \phantom{0}9.7 & \hspace{0.3 cm}  
   $-4.7 \pm 23.0$ & \textemdash& 22.0 & $< 7.2 $ \\  [1ex]
			%\hline
			$\overline{D}^{0} \rightarrow \overline{p}e^{+}$ & \phantom{0}9.6 &\hspace{0.4 cm} $7.1  \pm 9.0$ & 0.6& 23.0 & $< 7.6 $  \\  [1ex]
			%\hline
			$D^{0} \rightarrow p\mu^{-}$ & 10.7 & \hspace{0.5 cm}$11.0 \pm 23.0$ & 0.9& 17.1 & $<5.1$  \\  [1ex] 
			%\hline
			$\overline{D}^{0} \rightarrow p\mu^{-}$ & 10.7 &\hspace{0.3 cm}$-10.8 \pm 27.0$ & \textemdash& 21.8 & $<6.5 $  \\ [1ex] 
			%\hline
			$D^{0} \rightarrow \overline{p}\mu^{+}$ & 10.5 &\hspace{0.33 cm}  $-4.5 \pm 14.0$ & \textemdash& 21.1 & $<6.3$ \\ [1ex] 
			%\hline
			$\overline{D}^{0} \rightarrow \overline{p}\mu^{+}$ & 10.4 & \hspace{0.34 cm}$16.7 \pm 8.8$ & 1.6& 21.4 & $<6.5$  \\ [1ex]  
			\hline \hline
	\end{tabular}  }
	   \end{center}
    \end{table}

\begin{table}[htbp]
	\begin{center}
		 %\footnotesize
		%\label{syst}
   \caption{\label{tab:syst} Summary of systematic uncertainties (\%) due to normalization mode $D^{0} \rightarrow K^{-}\pi^{+}$ (Norm), proton ($p$) and lepton ($\ell$) identification, PDF used for signal extraction, fit bias from toy study (`-' denotes negligible bias) and tracking for $D^0 \to p \ell$ decay modes.}
		 %\resizebox{10cm}{!}{%
			\begin{tabular}{c  c  c  c c c  c c} 
				\hline \hline
				Decay mode & Norm & $p$ & $\ell$ & PDF & Fit Bias & Tracking &Total\\ 
    \hline \hline
    %\cline{2-7} \cline{10-11}
				%&Corr.&Err (\%)&Corr.&Err (\%)&Corr.&Err (\%)& & &Corr.&Sys (\%) \\ \hline
				$D^{0} \rightarrow pe^{-}$ &  1.8 & 0.3 & 1.6& $^{+4.7}_{-1.1}$ & - & 0.7  & $^{+5.3}_{-2.8}$\\ [1ex] %\hline
				$\overline{D}^{0} \rightarrow pe^{-}$ & 1.8 & 0.3 & 1.6&$^{+8.2}_{-9.2}$ & - & 0.7 & $^{+8.6}_{-9.5}$\\[1ex] % \hline
				$D^{0} \rightarrow \overline{p}e^{+}$ & 1.8 & 0.3 & 1.6&$^{\hspace{-0.1 cm}+11.1}_{\hspace{0.0 cm}-8.1}$ & - & 0.7 & $^{\hspace{-0.1 cm}+11.4}_{\hspace{0.0 cm}-8.5}$ \\ [1ex] %\hline
				$\overline{D}^{0} \rightarrow \overline{p}e^{+}$ & 1.8 & 0.3 & 1.6 &$^{+6.4}_{-2.0}$ & 1.7 & 0.7 & $^{+7.1}_{-3.6}$\\  [1ex] %\hline
				$D^{0} \rightarrow p\mu^{-}$ & 1.8 & 0.4 & 1.8&$^{+3.7}_{-7.8}$ & 2.2 & 0.7 & $^{+5.1}_{-8.5}$ \\  [1ex] %\hline 
				$\overline{D}^{0} \rightarrow p\mu^{-}$ & 1.8 & 0.3  &2.0&$^{\hspace{-0.05 cm}+21.3}_{\hspace{-0.05 cm}-12.3}$ & - & 0.7 & $^{\hspace{-0.05 cm}+21.5}_{\hspace{-0.05 cm}-12.6}$ \\ [1ex] %\hline
				$D^{0} \rightarrow \overline{p}\mu^{+}$ & 1.8 & 0.3 & 1.5 &$^{+21.7}_{-18.2}$ & - & 0.7 & $^{+21.8}_{-18.4}$\\ [1ex] %\hline
				$\overline{D}^{0} \rightarrow \overline{p}\mu^{+}$ & 1.8 & 0.3 & 1.6&${\hspace{0.1 cm}6.0}$ & 2.4 & 0.7 &  ${\hspace{0.1 cm}6.9}$ \\ 
				\hline		 \hline	
		\end{tabular}
              %}  
	\end{center}
 \end{table}

 The branching fractions ($\mathcal{B}$) for $D^0\to p \ell$ decay modes are measured using:
\begin{equation} \label{eq:2}
\mathcal{B}(D \rightarrow p\ell) = \frac{N_{p\ell}}{N_{K\pi}}\frac{\epsilon_{K\pi}}{\epsilon_{p\ell}} \mathcal{B}(D^{0} \rightarrow K^{-}\pi^{+}),
\end{equation}
\noindent
where $N_{p\ell}$ ($N_{K\pi}$) and $\epsilon_{p\ell}$ ($\epsilon_{K\pi}$) are the number of $D^{0}$ mesons and the efficiency for the signal (normalization) mode, respectively. The obtained signal yields, reconstruction efficiencies, upper limits on the signal yield and branching fractions for $D^0 \to p \ell$ decay modes are summarized in Table~\ref{tab:ul_data}. 

The statistical significance ($\mathcal{S}$) of the signal yield is evaluated using the likelihood ratio, $\mathcal{S}=\sqrt{-2\rm ln\,(\mathcal{L}_{0}/\mathcal{L}_{max})}$, where $\mathcal{L}_{\rm max}$ ($\mathcal{L}_{0}$) is the likelihood of the nominal fit (null hypothesis). The signal significances obtained for the $D^0 \to p \ell$ decay modes are summarized in Table~\ref{tab:ul_data}, after taking into account the systematic uncertainties. In the absence of a significant signal, an upper limit is calculated for each signal yield at 90\% confidence level using a frequentist technique~\cite{frequentist}. We generate toy experiments using the shape and background yield from the fitted PDF and vary the input signal yield. 
For each input value, we calculate the fraction of the ensemble that gives a fitted signal yield less than or equal to what is obtained in the data.
The upper limit on the signal yield is given at 90\% confidence level. The upper limit on the branching fraction is calculated using Eq.~\ref{eq:2} with $N_{p\ell}$ replaced by $N_{p\ell}^{UL}$. Systematic uncertainties are included in the upper limit calculations by smearing the signal yield with the systematic uncertainty. The resulting upper limits on the signal yields and branching fractions are listed in Table~\ref{tab:ul_data}.

Table~\ref{tab:syst} summarizes the systematic uncertainties in the measured branching fractions from various sources. The uncertainty on the track-finding efficiency is obtained using a control sample $D^{*+} \rightarrow D^{0}\pi_{s}^{+}, D^{0} \to \pi^{+} \pi^{-} K_{S}^{0}$, $K_{S}^{0} \rightarrow \pi^{+}\pi^{-}$ and is found to be 0.35\% per track. For the particle identification efficiencies, calibration factors are obtained using $D^{*+} \rightarrow D^{0}\pi^+_{s} (D^{0} \rightarrow K^{-}\pi^{+})$, $\gamma \gamma \to \ell \ell$, and $\Lambda \to p \pi^{-}$ control samples for pion, leptons, and protons, respectively to account for the differences between data and simulation. The resulting systematic uncertainties on the proton, pion, electron, and muon identification efficiencies lie within the range $(0.3-0.4)\%$, $(0.2 - 1.0)\%$, 1.6\%, and $(1.6-2.0)\%$, respectively, depending on the decay mode. The largest source of systematic uncertainty is the modeling of the $M_{D^{0}}$ and $\Delta M$ PDFs. The uncertainties on the PDF shapes are obtained by varying each of the fixed parameters by $\pm 1\sigma$ and the ratio of the differences in obtained and nominal signal yield to the latter is added in quadrature to calculate the total systematic uncertainty from PDF modeling. The fit procedures are validated in simulated MC samples. Small observed biases of $(1-2)\%$ are taken as systematic uncertainties. The uncertainties due to pion and kaon identification are $1.1\%$ and $0.8\%$, respectively, for the normalization mode $D^{0} \rightarrow K^{-}\pi^{+}$. The uncertainty on $\mathcal{B}(D^{0} \rightarrow K^{-}\pi^{+})$ is $0.03\%$. These uncertainties are taken into account to determine the upper limit on the branching fraction. The uncertainty on the $\pi_{s}$ efficiency cancels between the signal and normalization mode. All systematic uncertainties are added in quadrature to obtain the total systematic uncertainty. 

In summary, we have searched for the baryon and lepton number violating decays $D^{0} \rightarrow pe^{-}$, $\overline{D}^{0} \rightarrow pe^{-}$, $D^{0} \rightarrow \overline{p}e^{+}$, $\overline{D}^{0} \rightarrow \overline{p}e^{+}$, $D^{0} \rightarrow p\mu^{-}$, $\overline{D}^{0} \rightarrow p\mu^{-}$, $D^{0} \rightarrow \overline{p}\mu^{+}$, and $\overline{D}^{0} \rightarrow \overline{p}\mu^{+}$ by analyzing $921\,\mathrm{fb}^{-1}$ data collected at and 60 MeV below the $\Upsilon$(4S) resonance, and at the $\Upsilon$(5S) resonance by the Belle detector at KEKB. In the absence of any significant signal, we set an upper limit at 90\% confidence level for each signal decay mode. The corresponding limits on the branching fractions are determined to be (5 -- 8) $\times 10^{-7}$. The obtained upper limits are the most stringent to date. The limits on the $D \rightarrow p\mu$ modes are the first such results.

This work, based on data collected using the Belle detector, which was
operated until June 2010, was supported by 
the Ministry of Education, Culture, Sports, Science, and
Technology (MEXT) of Japan, the Japan Society for the 
Promotion of Science (JSPS), and the Tau-Lepton Physics 
Research Center of Nagoya University; 
the Australian Research Council including grants
DP210101900, % Urquijo
DP210102831, % Sevior
DE220100462, % Hsu
LE210100098, % Infrastructure
LE230100085; % Infrastructure
Austrian Federal Ministry of Education, Science and Research (FWF) and
FWF Austrian Science Fund No.~P~31361-N36;
National Key R\&D Program of China under Contract No.~2022YFA1601903,
National Natural Science Foundation of China and research grants
No.~11575017,
No.~11761141009, 
No.~11705209, 
No.~11975076, 
No.~12135005, 
No.~12150004, 
No.~12161141008, 
and
No.~12175041, 
and Shandong Provincial Natural Science Foundation Project ZR2022JQ02;
the Czech Science Foundation Grant No. 22-18469S;
Horizon 2020 ERC Advanced Grant No.~884719 and ERC Starting Grant No.~947006 ``InterLeptons'' (European Union);
the Carl Zeiss Foundation, the Deutsche Forschungsgemeinschaft, the
Excellence Cluster Universe, and the VolkswagenStiftung;
the Department of Atomic Energy (Project Identification No. RTI 4002), the Department of Science and Technology of India,
and the UPES (India) SEED finding programs Nos. UPES/R\&D-SEED-INFRA/17052023/01 and UPES/R\&D-SOE/20062022/06; 
the Istituto Nazionale di Fisica Nucleare of Italy; 
National Research Foundation (NRF) of Korea Grant
Nos.~2016R1\-D1A1B\-02012900, 2018R1\-A2B\-3003643,
2018R1\-A6A1A\-06024970, RS\-2022\-00197659,
2019R1\-I1A3A\-01058933, 2021R1\-A6A1A\-03043957,
2021R1\-F1A\-1060423, 2021R1\-F1A\-1064008, 2022R1\-A2C\-1003993;
Radiation Science Research Institute, Foreign Large-size Research Facility Application Supporting project, the Global Science Experimental Data Hub Center of the Korea Institute of Science and Technology Information and KREONET/GLORIAD;
the Polish Ministry of Science and Higher Education and 
the National Science Center;
the Ministry of Science and Higher Education of the Russian Federation, Agreement 14.W03.31.0026, % from 15.02.2018
and the HSE University Basic Research Program, Moscow; % from 15.04.2021
University of Tabuk research grants
S-1440-0321, S-0256-1438, and S-0280-1439 (Saudi Arabia);
the Slovenian Research Agency Grant Nos. J1-9124 and P1-0135;
Ikerbasque, Basque Foundation for Science, and the State Agency for Research
of the Spanish Ministry of Science and Innovation through Grant No. PID2022-136510NB-C33 (Spain);
the Swiss National Science Foundation; 
the Ministry of Education and the National Science and Technology Council of Taiwan;
and the United States Department of Energy and the National Science Foundation.
These acknowledgments are not to be interpreted as an endorsement of any
statement made by any of our institutes, funding agencies, governments, or
their representatives.
We thank the KEKB group for the excellent operation of the
accelerator; the KEK cryogenics group for the efficient
operation of the solenoid; and the KEK computer group and the Pacific Northwest National
Laboratory (PNNL) Environmental Molecular Sciences Laboratory (EMSL)
computing group for strong computing support; and the National
Institute of Informatics, and Science Information NETwork 6 (SINET6) for
valuable network support.

\end{document}